\documentclass[useAMS,usenatbib,usegraphicx]{mn2e}

\bibpunct{(}{)}{;}{a}{}{,}

\usepackage{graphicx}
\usepackage{psfig}
\usepackage{subfigure}
\title[Modelling the spectra of Jupiter]{Modelling the near-IR spectra of Jupiter using line-by-line methods}
\author[L. L. Kedziora-Chudczer and J. Bailey]{Lucyna Kedziora-Chudczer\thanks{E-mail:lkedzior@unsw.edu.au}
and Jeremy Bailey\\ 
School of Physics, University of New South Wales, Sydney, NSW 2052,  Australia}
\begin{document} 

\date{Accepted  Received ; in original form} 

\pagerange{\pageref{firstpage}--\pageref{lastpage}} \pubyear{2011}


\maketitle 

\label{firstpage}

\begin{abstract} 
We have obtained long-slit, infrared spectra of Jupiter with the Anglo Australian Telescope in the K and H
bands at a resolving power of 2260.  Using a line-by-line, radiative 
transfer model with the latest, improved spectral line data for methane and ammonia, 
we derive a model of the zonal characteristics in the atmosphere of this giant planet. 
We  fit our model to the spectra of the zones and
belts visible at 2.1 $\mu$m using different distributions of cloud opacities. 
The modeled spectra for each region match observations remarkably well at K band and in
low pressure regions at the H band. Our results for the upper deck cloud distribution are 
consistent with previous models 
(Banfield et al.1998) fitted to low resolution, grism spectra. The ability to obtain and model high resolution planetary spectra
in order to search for weakly absorbing atmospheric constituents can provide better constraints on the chemical composition of
planetary atmospheres.

\end{abstract}


\section{Introduction} 
The infrared spectrum of Jupiter is dominated by absorption features due to
methane (CH$_4$) as well as other gases such as ammonia (NH$_3$) and collision induced absorption
due to H$_2$ and He. Methane, however, has a very complex rovibrational spectrum and the available
spectrum line data are limited. Many models for Jupiter's spectrum have therefore been based on low
resolution absorption coefficients or k-distribution data, rather than on line-by-line techniques.
For example \citet{strong93} and \citet{irwin05} have presented k-distribution models for
methane absorption at a resolution of 10 cm$^{-1}$, and \citet{irwin99b} have presented similar
data for ammonia. Using data such as this it is possible to model spectra at relatively low
resolution such as those from Galileo-NIMS. However, many infrared spectrographs on ground-based
telescopes can provide much higher spectral resolution. Recently there have been significant
improvements in the available spectral line data for methane and ammonia \citep{albert09, rothman09,
yurchenko09, nikitin10, wang10}.

In this paper we present a set of observations of the spectrum of Jupiter at a spectral
resolution of $\sim$2260 and investigate our ability to model this data using line-by-line
methods. We use the radiative transfer model VSTAR (Versatile Software for Transfer of
Atmospheric Radiation), which is specifically designed
for this purpose \citep{bailey06}. It derives radiative transfer solutions for specified
wavelength ranges
in the transmission, emission or reflection spectrum of a planet. This software is designed to be
extremely versatile in the range of atmospheres it can be applied to. In the past we have used
it for the atmospheres of Earth and Venus \citep{bailey07, bailey08, bailey09} and recent additions to the 
system allow it to be used for the atmospheres of
solar-system planets, exoplanets, brown dwarfs and cool stars.

Methane is an important absorber, not just in Jupiter and the other solar system giant
planets, but also in brown-dwarfs (the presence of methane absorption is the defining
characteristic of a T-dwarf) and is likely to be important in many exoplanets. It has been
reported in the spectra of the transiting exoplanets \citep{swain08, swain09}. While hot methane
presents additional issues compared with the cool methane spectra required for Jupiter
\citep[e.g.][]{borysov02} the problem of accurately modelling the methane absorption spectrum is
common to a full understanding of all these objects. Thus Jupiter can be considered a test
object for our future ability to model the spectra of extrasolar planets, as well as an
interesting object in its own right.

Despite decades of research dedicated to understanding the structure and composition of Jupiter's
atmosphere, unanswered questions remain about the detailed distribution and transport of gases
over the surface of the most massive planet of the Solar system. The wealth of data across its
optical and infrared spectrum from ground-based telescopes \citep[e.g.][]{clark79, karkoschka94}, 
Pioneer 10 \citep{tomasko78}, Voyager \citep{conrath86}, Galileo
\citep{carlson96} and the Cassini spacecraft \citep{kunde04, bellucci04} provided direct measurements of
physical properties at different heights of the atmosphere, which can be used to constrain input
parameters in models. Spectral information is used not only to derive the atomic and molecular
abundances of the atmosphere given the physical properties such as temperature and pressure at
different heights, but also to study the distribution of aerosols and clouds, and dynamic 
characteristics, which include wind and temperature patterns due to global weather effects.

 Jupiter is covered with permanent layers of clouds, which are arranged into longitudinal,
relatively stable regions determined by atmospheric circulation. The bright and dark bands are
called zones and belts respectively, which are understood to correspond to up- and down-welling of
gas in the convective flow of the atmosphere. The most abundant constituents of Jupiter's atmosphere
are hydrogen, helium, methane and ammonia. Direct measurements of the scattering properties in Jupiter
atmosphere taken with  the Galileo Probe Nephelometer \citep{ragent98} suggest that the dense clouds
of condensates, mostly ammonia, form in the almost adiabatic troposphere above 1600 to 700~mb. In
addition the thinner, stratospheric hazes, which contribute to the overall atmospheric opacity, have
been detected at heights, which vary depending on the longitude of the planet. For example they
appear elevated in the polar regions, where hydrocarbons, such as C$_{2}$H$_{6}$ and C$_{2}$H$_{2}$
are suggested to exist as a product of photochemical and auroral discharge processes \citep{pryor91}. 
Another molecule considered in a composition of the stratosphere is photochemically
produced N$_{2}$H$_{4}$ \citep{strobel83}. The detailed composition of stratospheric haze remains
uncertain due to limitations of currently available data. A comprehensive review of Jupiter's cloud
structure was presented by \citet{irwin99a}. 

The observations described in this paper are long slit spectra at relatively high resolving power 
(R = $\lambda/\Delta\lambda\sim$ 2260) in the H and K band (Table~\ref{t1}). The data were obtained with the Infrared Imager and
Spectrograph 2 (IRIS2), \citep{tinney04} at
the Anglo-Australian 3.9 Telescope. The spectra were subdivided into latitudinal regions,
that closely resemble the pattern of zones and belts visible in the acquisition image observed with CH$_{4}$
filter at 1.673 $\mu$m (Figure~\ref{f1}). The details of  calibration technique are discussed in
Section~\ref{1}. We used the VSTAR package to develop models of the cloud structure for each region to
best fit the observed spectra. In Section~\ref{2} we describe the VSTAR modeling strategy to achieve this
goal, which includes the model parameters and the lists of spectral lines used for each constituent of
the atmosphere. The derived cloud opacities at different latitudinal locations are presented in
Section~\ref{3}, and followed by discussion in Section~\ref{4}.  

\section{Observations and Data Reduction} \label{1} 
\begin{figure}
\begin{center}
\psfig{file=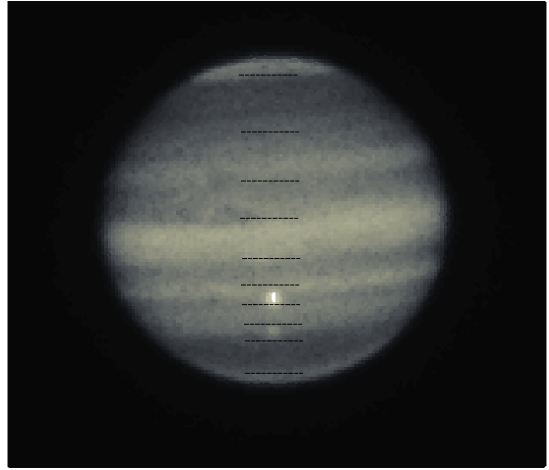,clip=,width=80mm}
\caption{The narrow-band, acquisition image of Jupiter taken in the CH$_{4}$ absorption region at 1.673 $\mu$m shows a
characteristic band structure of clouds in the atmosphere of the planet. 
The black lines mark the borders between regions, which correspond to zones and belts defined by the level of absorption visible
at 2.1 $\mu$m through a 1$^{\prime\prime}$ slit positioned vertically at the centre of the planetary disk. 
All observed spectra are derived as the averages over the number of pixels comprising each zone or belt.}\label{f1}
\end{center}
\end{figure}

The images and the IRIS2 spectra of Jupiter were
acquired on the 25th of May 2007 (Table~\ref{t1}). The long slit (7.7$^{\prime}$) was positioned across the meridian
of the planet. The width of the slit is 1$^{\prime\prime}$, which is equivalent to 2.2 pixels on the detector. The
spectra cover the H (1.62-1.8 $\mu$m) and K (2.0-2.4 $\mu$m) bands. The narrow-band image of the
whole planet in CH$_{4}$ (Figure~\ref{f1}) shows the characteristic band cloud structure. The
calibrated spectra were extracted and averaged over 11 regions denoted by names given
in Table~\ref{t2}. The region selection was based on the reflectivity profile averaged across the slit
over the section of the K band spectrum in the region of the collision induced H$_{2}$-H$_{2}$ absorption
between 2.08 and 2.13 $\mu$m.  The black horizontal lines overploted on the image in Figure~\ref{f1}
show the extent of each region. 

Data reduction was carried out using standard procedures of the Figaro package
\citep{shortridge95}.  The wavelength calibration was achieved by using a Xenon lamp. The spectra
were divided by spectroscopic flat fields in order to remove the non-uniform
response of the detector. The normalized flat-field images were formed by
subtracting the image fully illuminated by a quartz calibration lamp and the dark image taken with the
lamp switched off.  Sky subtraction was achieved by taking four exposures with two
spectra spatially offset in the same frame and subtracting two corresponding frames with offset spectra from each other.  

Spectral curvature was removed by resampling the frames using correction derived from spectra of
standard stars taken for each band. Bad pixels due to cosmic rays and known imperfections of the
detector were corrected by interpolation between their neighbours.

We used standard stars, which were close in angular separation to the planet. They were chosen to
match the solar spectral type as closely as possible. The Jupiter spectrum was divided by the standard
star spectrum to remove both telluric absorptions and solar features.  The resulting spectrum
corresponds to the reflectance from the planet. \citet{bailey07} have pointed out potential problems
with using standard stars for removal of telluric absorption for planetary spectra. However, in the
case of Jupiter the absorption features in the planet are due to different molecules from those in the
terrestrial atmosphere spectrum and these problems are largely avoided.

A true level of intensity can be derived by converting a number of counts,
$N^{c}$ in the spectra into the flux density derived from the calibrated flux density, F$_{st}$,
of the standard star.  We derive the radiance factor, I/F$_{Sun}$, where I is the reflected
intensity of the planet and $\pi$F$_{Sun}$  is the solar flux incident on the planet from:

\begin{equation}
\frac{I}{F_{Sun}} = \pi \frac{4D^{2}(N^{c}_{Jupiter})F_{st}}{\pi F_{Sun}(N^{c} _{st})(A_{slit})^{2}}
\end{equation}

The size of the slit, A$_{slit}$ is given in radians, and the distance of Jupiter, D from the Sun
is measured in astronomical units. The flux in each region is then scaled by the number of pixels, 
it encompasses.

\begin{table*}\begin{center} \caption{Jupiter observations on the 25 May
2007}\label{t1} \begin{tabular}{lcccccc} 
\hline Band  & Centre   &  Resolving power   & Mean dispersion &Start (UT) & Integration Time & Total \\
       & ($\mu$m) &      & (nm/pixel)      & (h)       & (s)              &(s) \\
\hline H    & 1.637    & 2270 & 0.341           & 14:40     & 12               & 48 \\ 
\hline K     & 2.249    & 2250 & 0.442           & 14:58     & 45               & 180 \\ 
\hline \end{tabular}
\end{center} \end{table*}

\begin{figure}
\centerline{\psfig{bbllx=25pt,bblly=140pt,bburx=585pt,bbury=700pt,file=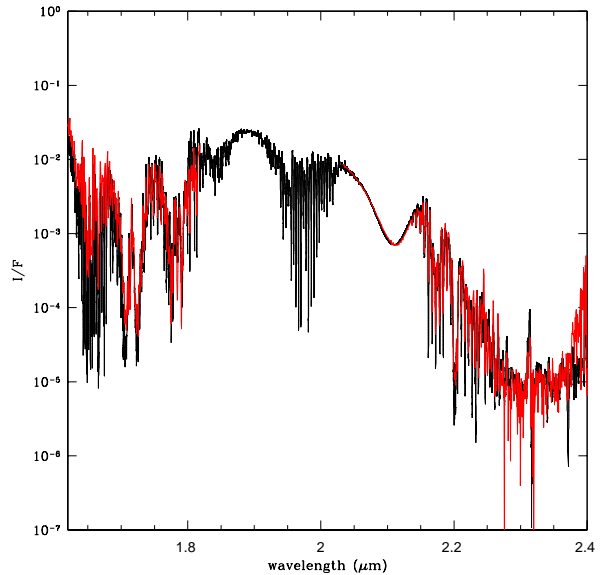,clip=.,width=80mm}}
\caption{The observed spectrum (grey or red in the online version) of Jupiter's equatorial region (EZ) at K and H band overlapped with the model (black),
which also cover the region between 1.85 and 2 $\mu$m characterized by the strong H$_{2}$O absorption in the Earth's atmosphere and
thus it is difficult to observe from the Earth.  
The main discrepancy between the model and observations appears at the low absorption region below 1.65 $\mu$m, which probes
atmosphere levels below 1 bar. In addition we find typically larger mismatch in intensity of the methane absorption bands at H
band than in K band, which is discussed in Section~\ref{3}.}
\label{f2}
\end{figure}

The typical spectrum and our VSTAR model in the equatorial region of Jupiter are presented
in Figure~\ref{f2} in the wavelength range between 1.62 and 2.4 microns.  As expected the
methane absorption clearly dominates in this spectral range.  The H$_{2}$-H$_{2}$,
collision induced smooth absorption peaks at 2.12 $\mu$m, which appears to be very shallow  in
the polar region spectra. This reflects the contributions of increasingly slant paths
through the atmosphere, which probe vertically higher regions with lower density of H$_{2}$.

The moon, Europa was transiting in front of the planet during the observations and its
reflected light was present in the slit while the spectra were taken. Its effect on the
spectra  of the Southern Equatorial Zone (SEZ) and Southern Belt (SB) is clearly be visible
in Figure~\ref{f5} showing the increased reflectivity in the regions of the largest methane
absorption.  Europa is thought to have the large amount of pure water ice on its surface
(see the spectra in \citet{calvin95}). The highly reflective shoulders of water bands
observed in the spectra around 1.8 and 2.2 $\mu$m contribute  to the combined high albedo
of the planet and the moon in the regions where otherwise the planetary methane absorption
is the highest.  

\begin{table}
\centering
\caption{List of selected regions of Jupiter}\label{t2}
\begin{tabular}{ll}
\hline  & Region\\
\hline 1 & North Pole (NP)  \\
\hline 2 & Northern Upper Belt (NUB)  \\
\hline 3 &Northern Zone (NZ) \\
\hline 4 & Northern  Equatorial Belt (NEB)  \\
\hline 5 &Equatorial Zone (EZ) \\
\hline 6 &Equatorial Belt (EB)  \\
\hline 7 & Southern Equatorial Zone (SEZ) $^a$ \\
\hline 8 & Southern Belt (SB)  $^a$ \\
\hline 9 & Southern Zone (SZ)  $^b$ \\
\hline 10 &Southern Upper Belt (SUB)  \\
\hline 11 & South Pole (SP)  \\

\hline
\end{tabular}
\medskip\\
$^a$These spectra are affected by a contribution of light reflected by Europa.
$^b$A spectrum of the atmospheric region with a bright 'white' spot.\\
\end{table}

\section{The VSTAR modeling procedure} \label{2} The Versatile Software for Transfer of
Atmospheric Radiation (VSTAR) is a modular code written in Fortran, which solves the line-by-line, radiative
transfer problem for light in a stratified atmosphere with a defined chemical composition and
a specified size distribution of cloud particles. The code is composed of several packages (MOD, LIN,
RAY, PART and RT) which contain routines called by a user written program to build a specific
atmospheric model.

The MOD package defines a two dimensional grid, which stores the values of parameters 
for the atmospheric layers in one dimension and the array of spectral points at which the model is
calculated in the second dimension.   The LIN package reads spectral lines from the line databases,
calculate line profiles and sets the optical depth for each layer of the atmosphere. The RAY package
calculates Rayleigh scattering optical depth for the atmospheric layers. 
The  particle scattering from aerosols and clouds is
calculated with a Mie scattering code \citep{mishchenko02}. The PART package allows up to 20
different scattering particle modes with different size distributions, refractive indices and vertical
distributions to be included in the model. The workhorse of the VSTAR code is the RT package, which
performs the radiative transfer solution for each layer of the atmosphere using the
data set up by the previously described packages. Different radiative transfer solvers can be
selected but for the purpose of this work we use DISORT, the discrete ordinate solver described in
\citet{stanmes88}.

We model Jupiter's atmosphere using 36 vertical layers with specific physical
properties given by pressure, temperature, as well as chemical composition expressed in terms 
of gas mixing ratios. The P-T profile of the planet's atmosphere in the pressure range between 0.001-1 bar has been  
studied with infrared radiometry, radio occultations and direct sensing with the Galileo probe ASI experiment \citep{seiff98}. 
Most recent retrievals of atmospheric temperatures from the Cassini CIRS data published by \citet{simon06} show time-variable, 
zonal differences in P-T profiles, which are also present in the Voyager IRIS data. In the range of pressures probed with our model
the strongest zonal variations are of the order of a few Kelvin in the upper troposhere. 

\begin{figure}
\centerline{\psfig{bbllx=20pt,bblly=140pt,bburx=585pt,bbury=720pt,file=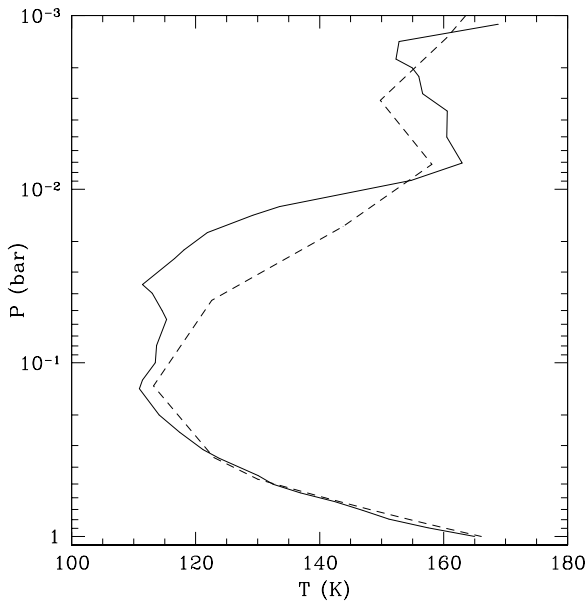,clip=.,width=80mm}}
\caption{Comparison between the temperature-pressure Voyager (solid line) and Galileo probe (dash line) profiles.}
\label{profile}
\end{figure}

In figure~\ref{profile} we plot the averaged profile from Voyager radio occultation experiments \citep{lindal92}, and the profile obtained
from the Galileo probe descending into the 5-$\mu$m hot spot region in the north equatorial belt \citep {seiff98}.
The notable differences in Voyager and Galileo profiles appear above about 300 mb in the upper troposphere, which is expected since 
both profiles were probing different locations on Jupiter surface at different times. The 5-$\mu$m hot spots are relatively dry, localized regions 
with a low level of cloud cover \citep{showman00}. We believe the Voyager occultation data are likely to be more robust in 
describing the average properties of zones and belts considered in our models.
We tested the effects of changed temperature corresponding to the profile differences observed in Voyager ingress, egress occultations data 
and Galileo probe measurements for the Equatorial belt, Equatorial zone and the North polar region. The typical example of a difference between models with Voyager 
and Galileo derived P-T profiles is shown in Figure~\ref{profile}. The differences are highest in the spectral ranges corresponding to the tropospheric pressures. 
However they are of the order of magnitude lower than the typical residuals of our fits to data, which are dominated by the opacities of different vertical 
regions in the atmosphere. 

Therefore for simplicity we used a single P-T profile to model zonal composition of the atmosphere 
based on the Voyager radio occultation observations \citep{lindal81, lindal92}.

\begin{figure}
\centerline{\psfig{bbllx=22pt,bblly=215pt,bburx=528pt,bbury=662pt,file=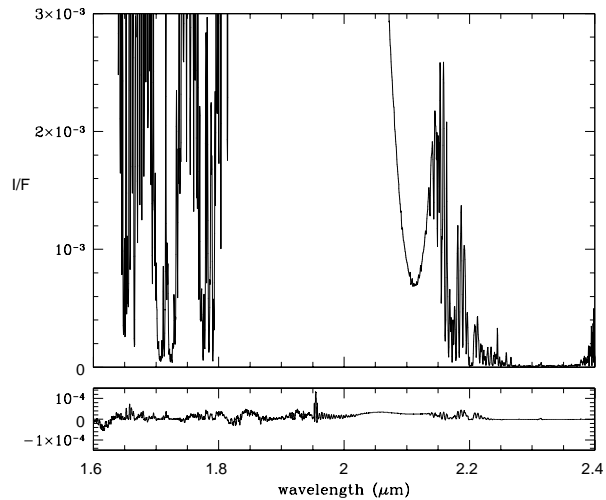,clip=.,width=80mm}}
\caption{The observed spectrum of Jupiter's equatorial region (EZ) at K and H band is shown with the difference between two models (in the lower panel) obtained 
with the Galileo Probe and the Voyager radio occultation P-T profiles,as presented in figure~\ref{profile}.}
\label{fpt}
\end{figure}

The chemistry of Jupiter's atmosphere is complex and involves a large number of atomic and
molecular species produced in photochemical processes \citep{atreya97}. Our model
considers only the major components of the atmosphere, which influence the shape of the observed
spectra in the near infrared region. The most abundant atmospheric gases are H$_{2}$ and He
that produce smooth, collision-induced absorption features around 2.12 $\mu$m. The other
molecules, which absorb in the infrared spectral region, are methane (CH$_{4}$) and ammonia
(NH$_{3}$). Methane is the most important absorber in the H and K band spectra. The
mixing ratio of CH$_{4}$ adopted in our model is $2.3\times10^{-3}$, which is consistent
with the latest in situ measurements with the Galileo Probe Mass Spectrometer (GPMS) \citep{wong04}. 
The mixing ratio of methane is relatively constant at pressures higher than 1 mbar \citep{moses05}.

\subsection{The Methane Spectrum}
\label{sec31}
A tetrahedral molecule like methane has four vibrational modes, two bending modes described
by quantum numbers $\nu_1$ and $\nu_3$ and two stretching modes $\nu_2$ and $\nu_4$, so that
each vibrational level can be described by its four vibrational quantum numbers $(\nu_1,
\nu_2, \nu_3, \nu_4)$. The frequencies of the two stretching modes are both about 1500
cm$^{-1}$ and the two bending modes both about 3000 cm$^{-1}$. These coincidences result in a
complex series of interacting vibrational states repeating at intervals of $\sim$1500
cm$^{-1}$ known as polyads. Polyads are generally designated by P$_n$ where n is the polyad
number defined in terms of the vibrational quantum numbers by

\begin{equation}
n = 2(\nu_1 + \nu_3) + \nu_2 + \nu_4
\end{equation}

P$_0$ denotes the ground state. The first polyad P$_1$ is also known as the Dyad since it
contains the two vibrational levels $(0, 1, 0, 0)$ and $(0, 0, 0, 1)$. The P$_2$ polyad (or
Pentad) contains 5 vibrational levels $(1, 0, 0, 0)$, $(0, 2, 0, 0)$, $(0, 0, 1, 0)$, $(0, 0,
0, 2)$ and $(0, 1, 0, 1)$. Similarly P$_3$ is the Octad (8 levels) and P$_4$ is the Tetradecad
(14 levels). The number of levels grows rapidly for higher polyad numbers and results in
complex systems of interacting bands which are increasingly difficult to model as the polyad
number grows.

A significant recent development is a new global analysis of the methane spectrum over the
wavenumber range 0--4800 cm$^{-1}$ (levels up to the Octad) reported by Albert et al. (2009).
The new model fits line positions and intensities much better than previous analyses of
individual polyads \citep[e.g.][]{hilico01}. The new model forms the basis of an improved
set of methane line parameters included in the 2008 edition of the HITRAN database \citep{rothman09}. 
It provides a good model of the low temperature methane spectrum for wavelengths
greater than about 2.1 $\mu$m.

However, at shorter wavelengths (higher polyads) HITRAN 2008 lists mostly empirical line parameters
from \citet{brown05}. These are measured line intensities and positions at room
temperature, and mostly lack quantum identifications and lower-state energies. The lack of a
lower-state energy means that line intensities cannot be reliably determined for temperatures
other than those of the original measurement. These lines are therefore not useful for
modelling the spectra of the giant planets. However, some lines in the strongest part of the
P$_4$ band system have empirically determined lower state energies derived from measuring the
lines at two or more different temperatures. These lines, with data from \citet{margolis90}
and \citet{gao09} cover the wavenumber range from 5500--6180 cm$^{-1}$.

For this work, we therefore constructed a methane line list as follows. We used line data
from HITRAN 2008 for all wavenumbers up to 4800 cm$^{-1}$. We also used HITRAN 2008 for the
region from 5500--5550 cm$^{-1}$, where  empirical
lower state energies are available. For the wavenumber range from 4700--5500 cm$^{-1}$, we used 
line parameters calculated using the Spherical Top
Data System \citep[STDS][]{wenger98}. This range overlaps slightly with
the range from HITRAN, but the lines from HITRAN are all in the P$_3$ polyad system in this
region, whereas the STDS lines are all from the P$_4$ polyad, so there is no duplication. 
While the STDS effective hamiltonian model for the Tetradecad region
is a preliminary one and does not give line positions that accurately match data, it does
provide an adequate description of the overall band shape, and provides the lower state
energies that are missing from HITRAN in these regions. For the range 5550--6236 cm$^{-1}$
we used line data from \citet{nikitin10} supplemented with low temperature empirical
data from \citet{wang10}.
We have not attempted to model 
wavelengths shorter than 1.6 $\mu$m. However, new empirical methane line data for these wavelength regions are 
starting to become available \citep{campargue10a,campargue10b,sciamma09}, which may enable our modelling to be extended to shorter
wavelengths.

The line width data provided in HITRAN are for broadening in air. Data on broadening of methane
lines in the $\nu_3$ band in a variety of broadening gases by \citet{pine92}, and \citet{pine03} shows that line
widths broadened by H$_2$ are about 2\% higher than those broadened by N$_2$, those in O$_2$
are about 6\% less, and those in He are about 35\% lower. This suggests that line widths in an
H$_2$/He atmosphere should be very similar to those in an N$_2$/O$_2$ atmosphere. We have
therefore used the air broadening data from HITRAN to determine the line widths in this work.
Where the line data are taken from sources without broadening parameters, we used the default line half width value of
$\gamma$ = 0.06 cm$^{-1}$ atm$^{-1}$ and a temperature 
exponent N = 0.85 as recommended by \citet{nikitin10}.

\subsection{Other Absorbers}

The atmospheric abundance of NH$_{3}$ at high pressures was measured directly with GPMS and
from the attenuation of radio signal sent between the probe and orbiter \citep{folkner98}. 
The mixing ratio appears to decrease from 10$^{-4}$ at pressures above 2 bar to
10$^{-6}$ below 1 bar.  Ammonia is known to exist in gaseous form and as condensed clouds
in Jupiter's troposphere up to pressures of  0.1 bar \citep{kunde82}, but it is
rapidly depleted in the stratosphere, due to UV photolysis. 

The mixing ratios of NH$_{3}$ adopted in our model are constant ($2\times 10^{-4}$) in
layers between 1 and 0.5 bar. They decrease to $2\times 10^{-7}$ at the 100 mbar along the saturated vapour pressure curve.
The resulting profile is consistent with mixing ratios derived from the radio data in \citet{sault04} above 0.5 bar.
They note small differences between the Equatorial Zone and North Equatorial Belt profiles,
which we also tested in our models. As a result we observe changes in the absorption spectrum between 
1.8 and 2 $\mu$m, where the $\nu_1 + \nu_4$ and $\nu_3 + \nu_4$ bands of NH$_{3}$ are present. 
We found that removing ammonia from the upper atmosphere ($ < 0.5$ bar), which is most likely the case in down-welling regions, 
like belts and hot spots, does not have a significant effect on the K and H band spectra 
which are dominated by CH$_{4}$ absorption.

Absorption of ammonia was derived from the new list of Yurchenko et al. (2009) which contains
over 3 million lines and is considerably more complete than the line parameters in HITRAN
2008, including improved shorter wavelength coverage.

The bulk of Jupiter's atmosphere consists of H$_{2}$ and He. Rotovibrational absorption
bands due to collisions between pairs of H$_{2}$-H$_{2}$ and  H$_{2}$-He exist in the near
infrared region under consideration, where they form a smooth feature centred around
2.1$\mu$m. Our model includes absorption coefficients for H$_{2}$-H$_{2}$ and
H$_{2}$-He pairs at low temperatures, derived by \citet{borysow02} and \citet{borysow89}
respectively. 

It is worth noting that the 2.1$\mu$m collision induced absorption (CIA) depends also on the
para-ortho ratio of hydrogen spin isomers in the atmosphere of Jupiter. Our current models
assume the equilibrium ratio at all temperatures.
The Voyager IRIS data was used by \citet{conrath98} to demonstrate that the distribution of that ratio varies in the 
different zonal and vertical regions of the planet. The deviation from so called 'normal', 1:3, room temperature, equilibrium value
is relatively low for Jupiter, but it can produce a noticable effect on the shape of the 2.1$\mu$m feature in the spectrum. 

Our intention is to extend the capability of the VSTAR package in the near future 
to allow differences of the para-ortho H ratio as a function of atmospheric height, which will allow more 
accurate fit of the CIA absorption in atmospheres of cold giant planets,
where the observed para-ortho ratios deviate even further from the normal value.

\subsection{Scattering}

In our model Rayleigh scattering from molecules and particle scattering from
aerosols suspended in the atmosphere was considered. The Rayleigh scattering at wavelengths
above 1 $\mu$m is much reduced in comparison with scattering from clouds. The Rayleigh cross
sections \citep{liou02} were calculated for scattering in a mixture of H$_{2}$  and He, the
dominant  components of the atmosphere.

The vertical structure of aerosols in Jupiter's atmosphere, derived directly from Galileo
observations was surprisingly different from theoretical predictions \citep{atreya97},
which can be attributed to the unique region of the Probe entry. The Galileo Solid State
Imager (SSI) observations covered more extended regions over belts and zones as described
in \citet{banfield98}. The SSI images taken close to the limb of Jupiter gave
good information about the cloud opacity in the stratosphere. 

\citet{banfield98} used a retrieval technique \citep{banfield96} to obtain the
scatterer density with altitude in the same spectral region as our observations. The upper 
troposphere between 700 and 100 mbar, and stratosphere at pressures lower than 100 mbar  
defined by Banfield et al. are within the spectral sensitivity of the H and K bands. The results of
the Banfield et al. retrieval  of particle distribution from the Galileo imaging suggest two
enhancements in cloud opacity corresponding to the upper haze  between 200 up to 20 mbar,
and at the base of the tropospheric, condensation cloud at about  850 mbar.  In equatorial
regions, the upper haze is elevated to even lower pressures.  

The layer of smaller particles is believed to form a relatively uniform top of the
tropospheric cloud,  which contains larger particles of presumably NH$_{3}$ ice and other
condensates. Observations of limb  darkening in the ultraviolet \citep{tomasko86}
constrain the size of the stratospheric haze to between 0.2 and 0.5 $\mu$m. To model
the clouds we chose a simple two-component distribution of mean particle sizes, 0.3$\mu$m for
the stratosphere and the conservative value of 1.3 $\mu$m size species for the upper 
tropospheric cloud. The combined observations from NIMS and NIR identify the lower
tropospheric cloud as a mixture of two types of particles with 50 and 0.45 $\mu$m size,
consistent with NH$_{4}$SH molecule \citep{irwin99a}. The upper troposphere, which our data is
sensitive to, is believed to consists of 0.75  $\mu$m size ammonia particles. 

The scattering properties of these clouds are calculated using Lorentz-Mie theory for a
power-law particle size distribution.  The code calculates the asymmetry factor,
extinction and scattering coefficient, which are required in the radiative transfer
calculations. The refractive index of the cloud particles is approximated by the
refractive index of ammonia ice given by \citet{martonchik84} for a wide range of
wavelengths.

In our model the vertical structure of the particle clouds is varied through the input list
of optical depths for each layer of the atmosphere at the reference wavelength of
1.5 $\mu$m. Finally the  boundary conditions for the model are given. Solar illumination
at a given beam angle is provided using data from \citet{kurucz09} and a non-reflecting
surface is specified at the base of the atmosphere. The set of line data, particle
properties and their distribution form inputs for the DISORT radiation transfer equation
solver, which returns the fluxes and radiances at the top of the atmosphere for each
wavelength specified in the grid. The models were calculated at a spectral resolution of 0.01 cm$^{-1}$,
sufficient to resolve the spectral structure, and then convolved with an instrumental point spread
function for comparison with the observed spectra.

\section{Model of the bands in the atmosphere of Jupiter}
\label{sec4}
To show the functionality of the VSTAR code, we used the most recent observational data described
in the previous section, which provided the input parameters for characterisation of
Jupiter's atmosphere. We tested the sensitivity of our models to changes in these parameters
and confirmed that spectral features are dominated by methane absorption with the addition
of clouds above the 800 mbar level. 

A simple model of the longitudinal belts and zones in the atmosphere of Jupiter was
derived by manually varying the amount and vertical distribution of cloud opacity at different heights, while the
chemical composition of the atmosphere was left constant across the surface of the planet.
Since this is clearly not practical to explore full range of opacities at all levels of the atmosphere, 
we resorted to testing different combinations of opacities in the iterative way and fine-tuned them until a reasonable fit, 
with the smallest scatter in residual differences between spectra and models was achieved for both spectral bands H and K.

\begin{figure}
\begin{center}
\psfig{bbllx=28pt,bblly=160pt,bburx=586pt,bbury=650pt,file=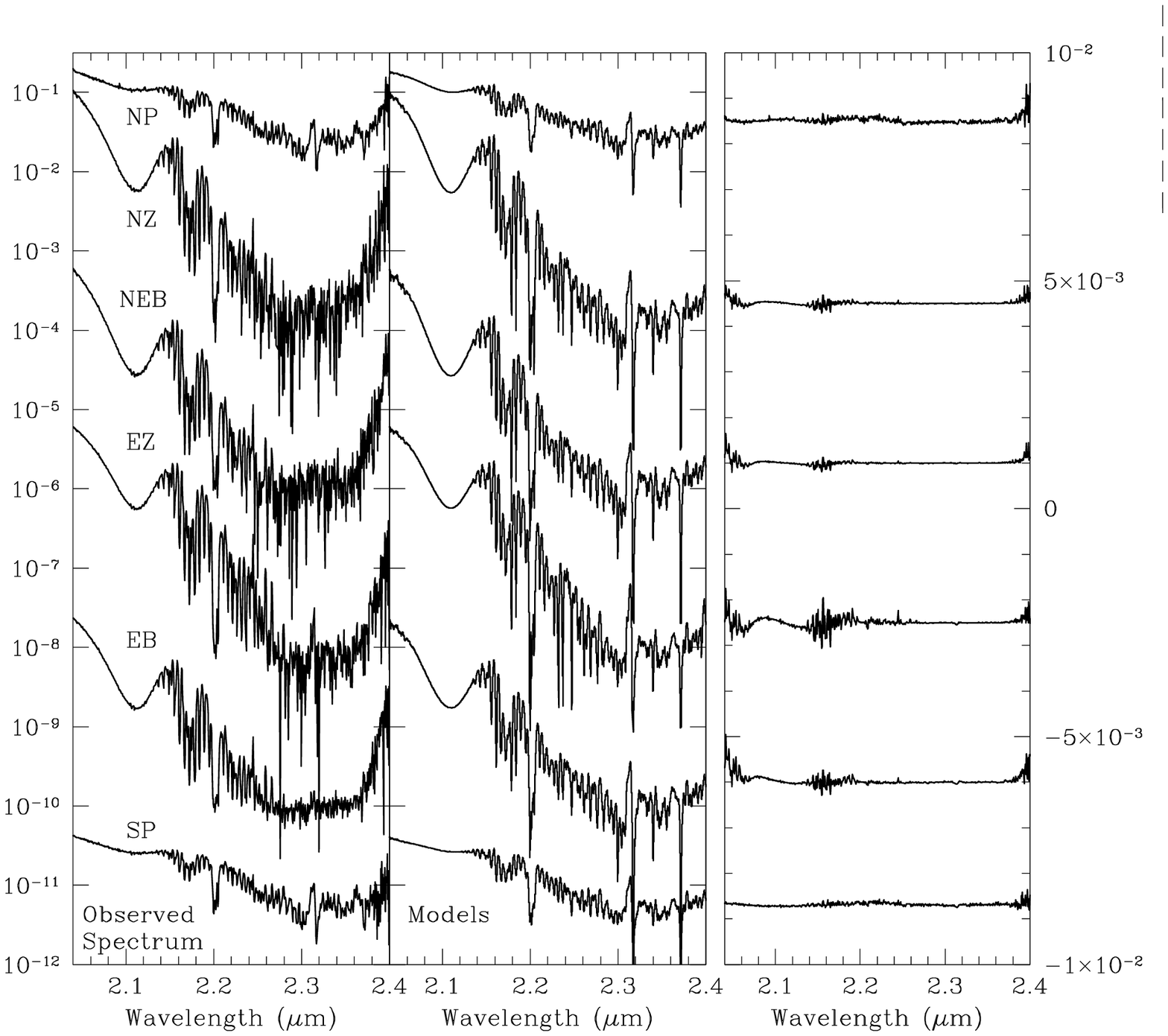,clip=,height=82mm,width=82mm}

\psfig{bbllx=28pt,bblly=160pt,bburx=586pt,bbury=650pt,file=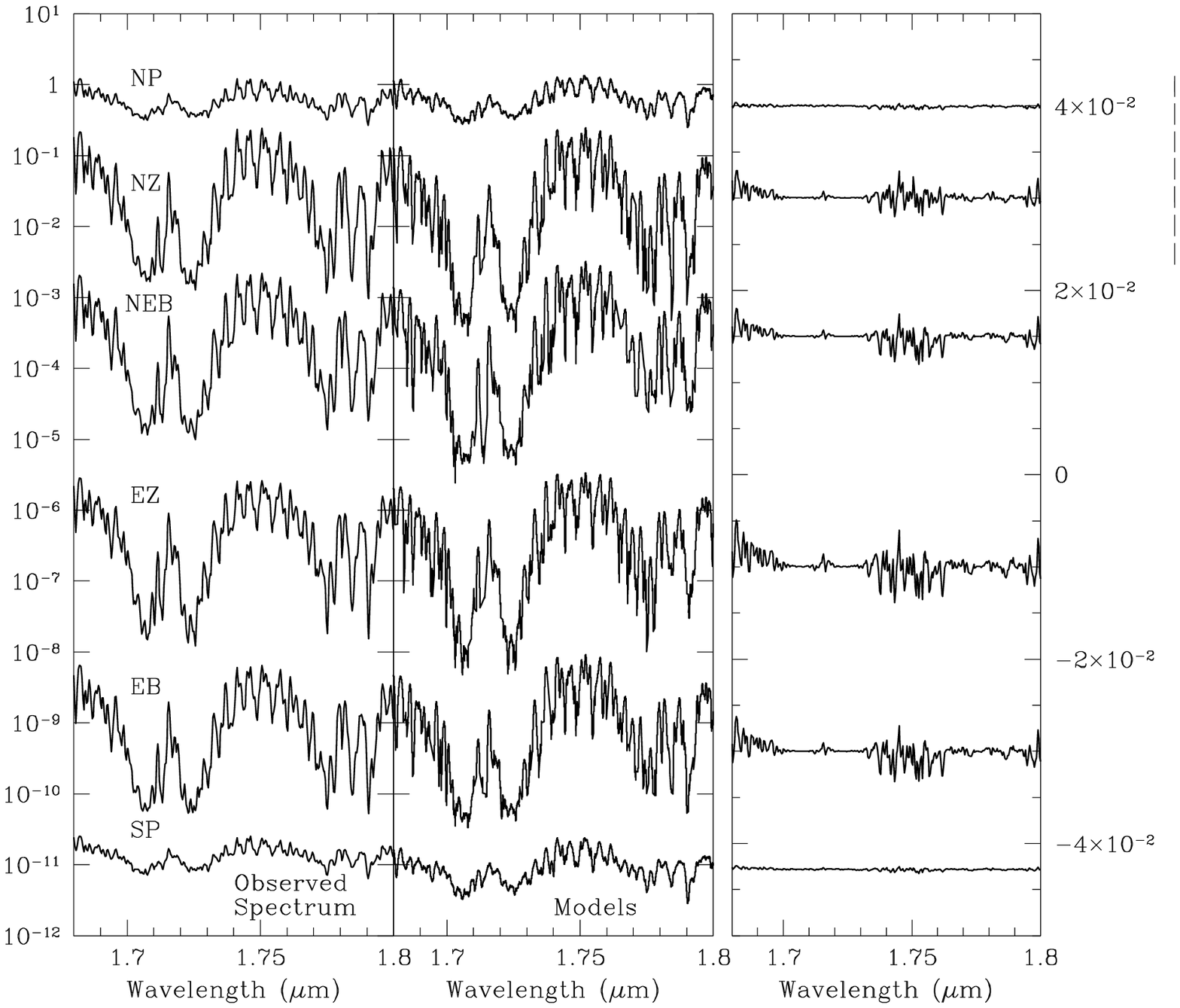,clip=,height=82mm,width=82mm}
\caption{In the upper panels the observed spectra of the selected zones and belts in the K band (on the logarithmic intensity scale) 
are compared with the corresponding VSTAR models derived for these regions. 
In the lower panel the observed spectra in the H band are compared with the VSTAR models. The panels on the right (top and bottom)
demonstrate the residuals obtained by taking the difference between the modelled spectra and corresponding data. }\label{f3}
\end{center}
\end{figure}
 
Figure~\ref{f3} shows the line-by-line models for the selected bands and zones with a resolution
that matches the observed spectra. Our models mirror most spectral features  remarkably accurately
considering that the average spectrum is a composite of sub-regions with possibly non-uniform cloud
composition as evident in high resolution images from the Galileo SSI probes
(Simon-Miller et al. 2001). We found that the most significant differences between the models and the data occurred in the equatorial zones, 
in spectral regions that correspond to the upper layers of the troposphere. 

The most recognizable features of the K band spectrum are the
H$_{2}$-H$_{2}$ collision-induced broad absorption centered at 2.11 $\mu$m and stronger absorption
in the 2.2 to 2.4 $\mu$m region due to methane bands. The narrower CH$_4$ absorption features
are due to the Q branches
($\Delta J =0$) of $\nu_{2}+\nu_{3}$ at 2.2 $\mu$m, $\nu_{3}+\nu_{4}$ at 2.32 $\mu$m
and  $\nu_{1}+\nu_{4}$ at 2.37 $\mu$m. The latter two bands are not always clearly visible in the observed
spectra due to low signal-to-noise in the high methane absorption region and a decreased sensitivity 
at the edge of K band filter.

Spectra of the polar regions show the broad
methane, and H$_{2}$-H$_{2}$ collision induced absorption much reduced as compared with the equatorial
zones and belts. This can be understood by noting that the spectra of the polar regions sample
relatively longer path-lengths through the higher layers of the atmosphere, which unlike the deep
layers contribute little to the H$_{2}$-H$_{2}$ absorption. 

\begin{table} \begin{center} \caption{Total cloud optical depths in zones and belts }\label{t3}
\begin{tabular}{ccccc} \hline 
Particles& 1.3$\mu$m & &0.3$\mu$m & \\ \hline  
Zones &Extent & $\tau$ & Extent & $\tau$\\ & (mb) & & (mb) & \\ \hline 
NP  &    563-631 &1.50 &56-200&  0.0100\\ 
& & &3-7 &0.0030\\ \hline 
NUB&316-562 &0.52 &125-200 &  0.0003\\ 
& & & 25-40 & 0.0015\\ 
& & & 7-9 & 0.0001\\ \hline 
NZ &316-631 &1.72 & 22-25&0.0010  \\ \hline 
NEB&355-751 &5.70 & 25-35 & 0.0018 \\ \hline
EZ& 251-631 & 5.72 & 18-22 & 0.0023\\ \hline 
EB & 316-708 & 5.09 &25-35 &0.0070 \\ \hline 
SEZ$^a$ &447-708 &3.10 &39-50 &0.0720\\  \hline 
SB$^a$& 355-708& 6.06 & 141-251 &  $0.0188 $ \\
& & &40-50&0.0090\\  \hline 
SZ$^b$ &446-631 &2.25 & 100-355 & $0.0295$ \\ 
& & & 25-35 & 0.0060\\ 
& & & 3-5 & 0.0010\\ \hline 
SUB& 316-751& 1.29 &141-316 & 0.0024 \\ 
&  & & 25-40 & 0.0080\\ \hline 
SP&251-708 &2.20 & 22-251& 0.0038 \\ 
& & & 2-4 &0.0012\\ \hline \end{tabular} \medskip\\
$^a$The spectra affected by a contribution of light reflected by Europa, which boosts reflection in the otherwise most highly
absorbing methane bands regions in K and H bands. This additional reflection was not included in the modelling.  
$^b$A spectrum of the atmospheric region with a bright, 'white' spot, which can be modelled as the relatively 
thick stratospheric cloud extending upwards to 100 mbar. \end{center}
\end{table}

The models are constrained by a simultaneous fit of absorption levels for the broad and
narrow band features of the spectra in both bands H and K.  The lower panel of
Figure~\ref{f3} shows the spectra and models of the H band. The narrow spectral lines in
the models tend to agree better with the observed spectrum than the broad shape and the
intensity of absorption.  We attribute this to the less accurate models of methane line
parameters in this spectral region as discussed in section~\ref{sec31}. The accuracy of the
wavelength fit is limited mostly by the spectral calibration of the data, dominated by the
correction of spectral curvature in the IRIS2 grism system. The wavelength of clearly
identified lines in the data  and models agrees within $5\times 10^{-4} \mu$m. Most
discrepancies in models and observed spectra are due to the differences in intensity of a
particular line or line series in methane absorption. 

The Jupiter K band spectrum is defined by the opacities distributed between about 700 and  2 mbar.
The shape of collision induced H$_{2}$-H$_{2}$ absorption, which samples deep layers of the
atmosphere is highly sensitive to the cloud optical depth in the Jupiter's troposphere, while the
level of broad methane absorption beyond 2.2 $\mu$m varies dramatically in response to changes in
the stratospheric cloud optical depth.  The H band spectral region samples the higher levels of the troposphere. 

Table~\ref{t3} shows the total optical depths of the tropospheric (1.3 $\mu$m particles) and
stratospheric aerosols (0.3 $\mu$m particles), which gave the best fit for each spectrum. 
All zones and belts show differences in the vertical distribution and the strength of
opacity. Most notably the atmosphere appears to be relatively clear between 200 and 40 mbar
in the northern and equatorial regions, which is consistent with conclusions in \citet{banfield98}, 
who find a similar clearing around 100 mbar in their models in the same spectral
region. We find that no trial models with added cloud opacity in this pressure range were able to
reproduce the observed spectra. 

We find that the clouds tend to lie deeper in belts than in zones. The tropospheric cloud
seems to extend up to 250 mb in the equatorial region with the highest total opacity of the
order of $\tau=5$, while the stratospheric haze is very thin with the opacities below $\tau
\sim$ 0.005. Our models suggest that this haze may be slightly thicker in the
southern hemisphere of the planet. However the SEZ and SB regions are
affected by the spectrum of Europa, while the spectrum of the SZ zone may reflect a local cloud pattern.

The image of Jupiter (Figure~\ref{f1}) shows a rather large white spot in this region, which is subtended by a slit. 
The high resolution images taken by the New Horizons LEISA imager at
1.53 and 1.88 $\mu$m in February 2007\footnote{On-line images available from the NASA/Johns Hopkins University Applied Physics
Laboratory/Southwest Research Institute ({\it http://photojournal.jpl.nasa.gov/catalog/PIA09251})} show 
similar weather features in the region
of SZ and SUB. This spot is most likely a storm system such as the anticyclonic white oval modelled
in Simon-Miller et al. (2001). They note that fresh material is upwelled in such high-pressure
systems, which forms slightly higher clouds. Our model of the SZ white spot requires a relatively
thick cloud at the 100-251 mb pressures with the small size particles, such as used for the stratospheric layers, 
which provided better fit to the observed spectra than any larger particles.

The zones clear of Europa light show higher stratospheric cloud opacities, which are distributed more
thinly than clouds in nearby belts. The stratosphere in the high latitude northern (NUB) zone is more
spread-out but relatively  thin compared with the equatorial regions. The stratosphere above the
southern upper (SUB) belt is also vertically extended, but in contrast to the NUB, has higher opacity. 
Previous models \citep{west86, simon01} suggested that downwelling of the haze
in the belt  regions could produce thinner cloud at high pressures. 
Our models show unambiguously increased opacity of high stratospheric layers in polar regions, which is consistent 
with the previous studies of Jupiter's atmosphere \citep{pryor91}.    

\begin{figure}
\begin{center}
\psfig{bbllx=22pt,bblly=215pt,bburx=575pt,bbury=702pt,file=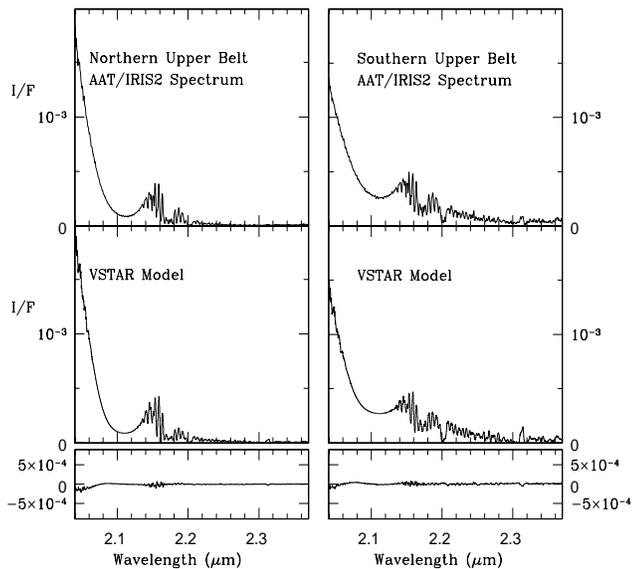,clip=,height=80mm,width=85mm}
\psfig{bbllx=22pt,bblly=215pt,bburx=575pt,bbury=702pt,file=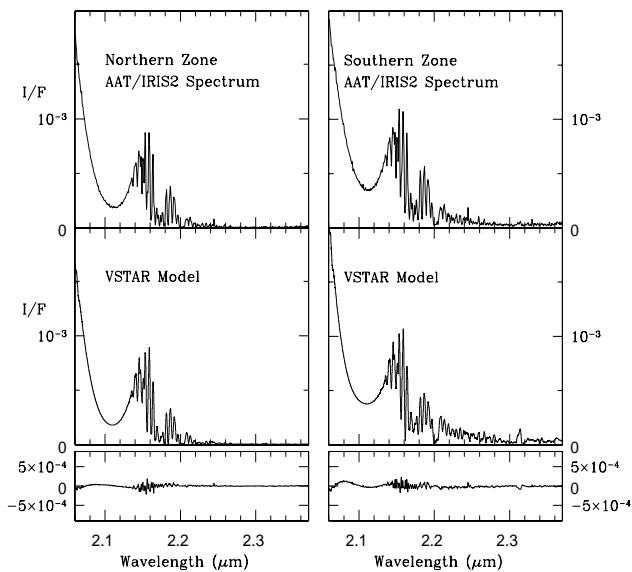,clip=,height=80mm,width=85mm}
\caption{The observed spectra, models and the difference between them (on the linear intensity scale) are shown for the northern and southern, high latitude zones (NZ and SZ) and belts (NUB and SUB). 
Although the spectra of the northen regions show typically stronger absorption visible in 
both H$_{2}$-H$_{2}$ feature and the wavelength region above 2.2 $\mu$m, this differences can be explained in terms of 'local' weather effects
 as discussed in Section~\ref{3}. }\label{f4}
\end{center}
\end{figure}

\begin{figure}
\begin{center}
\psfig{bbllx=22pt,bblly=215pt,bburx=575pt,bbury=702pt,file=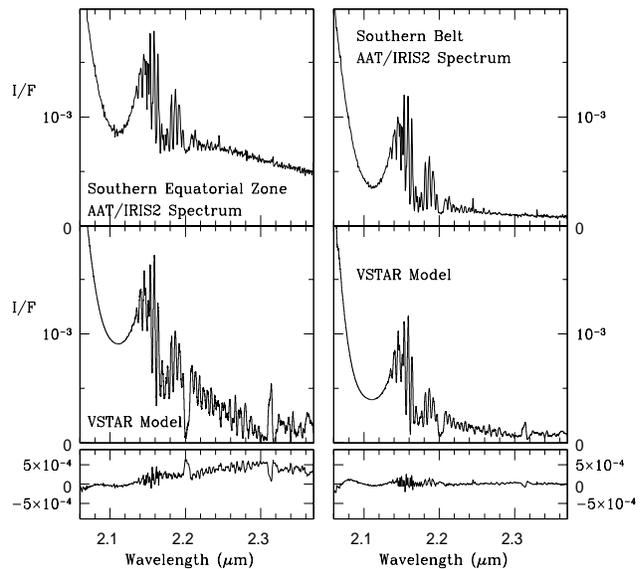,clip=,height=80mm,width=85mm}
\caption{The spectra of zones affected by opacity of Europa are shown compared with the simple models, 
which neglect the scattering contribution from this moon. The differences between spectra and models (bottom panels) show this clearly for spectral regions above 2.2$\mu$m .}\label{f5}
\end{center}
\end{figure}

In Figure~\ref{f4} the models of remaining regions are presented. Northern and southern
zones (NZ and SZ) display a slightly different shape of the collisional absorption, which
suggests the existence of more compressed, optically thicker tropospheric cloud in SZ
than the cloud in the NZ.  The SEZ and SB spectra (Figure~\ref{f5}) show the additional
opacity due to Europa's spectrum dominated by fine-grained water ice (Carlson
et al. 1996). The effects of this opacity are clearly visible  in the spectral region
beyond 2.15 $\mu$m, which is sensitive to the highest atmospheric levels in our model. 

\label{3}

\section{Discussion} \label{4} The line-by-line radiative transfer models for high resolution
 spectra from currently available  ground-based instruments serve as an important test of the
 consistency in our understanding of the composition and processes that govern planetary
 atmospheres.    

We achieved a good fit to the data assuming relatively simple physical structure and
chemical composition of Jupiter's atmosphere that was kept uniform across the whole surface of
the planet. The real atmosphere is certainly more complex. Lacking a detailed P-T profile for each
zone and belt we used a single profile of the atmosphere for all regions based on
radio occultation observations \citep{lindal92}.  This seems justified despite small differences
in atmospheric profiles observed by Voyager 1 infrared spectrometer (IRIS) and Cassini CIRS experiments \citep{simon06}
for different locations on the planet. 
It is argued, that because horizontal differences in observed brightness temperature are
small, the larger deviations in observed brightness temperature are probably due to varied opacity
in the clouds \citep{west86}. 

We also used a relatively simple representation of the chemical composition of the atmosphere, despite
the fact that the abundances of gases in different regions of the atmosphere may vary with height
due to evidently complex, planetary weather system. The upwelling in zones may cause the depletion of
volatiles due to condensation and precipitation. In the down-well regions the mixing ratios of such
volatiles may be increased in deeper levels of the atmosphere. While methane is well mixed within
the vertical region of our sensitivity, there have been zonal differences in mixing ratios profiles of ammonia
below 600 mbar suggested by \citet{sault04}.  Our models turned out to
be not sensitive to these differences due to combined effect of the low NH$_{3}$ absorption in our
spectral range and decreased sensitivity of our models to layers below $\sim 700$ mbar.  We find
that it is sufficient to consider only the major contributors to the spectral absorption at H and K
bands:  H$_{2}$, He, CH$_{4}$ and NH$_{3}$ and to modify the opacities of aerosols at different
heights. However our systematically worse fits in equatorial zones may reflect the unrecognized complexity in these regions, 
which is not described in sufficient detail by the initial parameters of our model, such as P-T profile and mixing 
ratios of the chemical components.  

The suggested constituents of clouds in the troposphere below 500 mbar are H$_{2}$O, NH$_{4}$SH and
NH$_{3}$.   Ammonia condensation occurs at the highest level at about 700 mbar at its saturation
point. The actual condensation level can be lowered or lifted due to differences in the molar
abundances of NH$_{3}$. \citet{west86} suggested that the cloud needs to have a mixed
composition of particles with sizes between 3 and 100 $\mu$m to explain high opacity observed at
45 $\mu$m. The larger particles are concentrated at the base of the cloud at pressures below our
sensitivity limit. Consistently our models are most sensitive to changes in the particle sizes below
3 $\mu$m. We find that the tropospheric cloud in belt regions occurs at typically higher pressures compared to
zones, which could be interpreted as the result of vigorous convective downwelling and upwelling process.

In contrast the stable haze of $\sim 1 \mu$m-size particles, mixture of ammonia ices and
residual chromophores responsible for the coloration of the clouds exists above the lower,
convective troposphere. In the warmer temperatures there is no condensation expected and
consequently no cloud formation. The origin of the stratospheric haze is thought to be due
to photochemical processes. 

This original picture of the layers above 700 mbar shown in \citet{west86} was
developed further by \citet{banfield98}, whose model includes a clearing of the haze at
about 100 mbar pressures. They explain it by the aerosol coagulation at this level of 
stratosphere, which is the cause of the increased, gravitational fallout of heavy
particles.  We observe such a clearing in models of most of the northern zones and belts of
the planet. Above the clearing we find the optically thin haze of reflective particles with
a predominantly lower sizes, than particles in the troposphere. This haze extends
to the low pressures of  2 mbar  in the polar regions. The chemical composition of the stratospheric 
haze is still debated. It is also no clear if the polar haze observed at lower pressures
has the same composition as the equatorial stratosphere.

\section{Summary}  We have used the new, line-by-line, radiative
transfer package, VSTAR, to model the near infrared spectra of Jupiter. The models were produced
for regions with different absorption properties implied by varied cloud patterns in the atmosphere
of the planet. With our improved database of methane absorption lines at low temperatures, we achieved
a good fit to the spectra obtained with the AAT/IRIS2. We were able to
reproduce conclusions suggested in previous studies about the atmospheric conditions at 
pressures below 1 bar, which probe the upper troposphere and stratosphere of the planet. 
We showed that ground-based spectroscopy with large telescopes can significantly enhance 
the results achieved from space. 
With continued improvements in methane line parameters, we will be able to extend our model
of the near IR spectrum of Jupiter and other giant planets and to cover full range of wavelengths 
between 1 and 2.5 $\mu$m with almost arbitrary resolution. This will make it possible to 
search for and measure weak spectral features due to trace constituents, that at present
cannot be picked out of the forest of methane lines. These include, for example, absorption due to
monodeuterated methane (CH$_3$D) with lines at $\sim$1.55 $\mu$m that can be used to measure the D/H ratio of the giant
planets with much higher accuracy than has been possible so far.
    
The most appealing feature of the VSTAR package is its versatility. Although slower than correlated-k techniques, 
it allows spectral modelling of a wide range of objects, from the cool brown dwarfs to cold planets and their moons.
These models can be calculated at any required spectral resolution and can therefore
be used in conjunction with data from instruments now available on 
large ground-based telescopes that can provide near-IR spectra at
resolving powers as high as 100,000. With the ever increasing speed of computing devices, it may be soon 
practical to use line-by-line techniques like VSTAR to achieve best fits to the high resolution spectra via automated retrival schemes.

The methane line list used in this work (in HITRAN 2004 format) is available for download from
http://www.phys.unsw.edu.au/\string~jbailey/ch4.html

\section*{Acknowledgments}
We acknowledge the support of the Anglo-Australian Observatory in scheduling the observations presented here in 
the IRIS2 service observing programme. We also thank the anonymous reviewers for constructive comments and valuable 
ideas towards the extension of this study. 


\label{lastpage}


\begin{thebibliography}{}
\bibitem[\protect\citeauthoryear{Albert et al.}{2009}]{albert09} Albert, S., Bauerecker, S., Boudon, V., Brown, L.R., 
Champion, J.-P., Loete, M., Nikitin, A. \& Quack, M., 2009, Chemical Physics, 356, 131.
\bibitem[\protect\citeauthoryear{Atreya et al.}{1997}]{atreya97} Atreya, S.K., Wong, M.H., 
Owen, T. C., Niemann, H. B. \& Mahaffy, P. R., 1997, {\it The Three Galileos: The Man, 
The Spacecraft, The Telescope}, Kluwer Academic Publishers, pp249--260
\bibitem[\protect\citeauthoryear{Bailey}{2006}]{bailey06} Bailey, J., 2006, in Forget, F et al. (eds) Proceedings Mars 
Atmosphere Modelling and Observations Workshop, Granada, 148, Laboratoire Meteorologie
Dynamique, Paris.
\bibitem[\protect\citeauthoryear{Bailey}{2009}]{bailey09} Bailey, J., 2009, Icarus, 201, 444.
\bibitem[\protect\citeauthoryear{Bailey, Simpson \& Crisp}{2007}]{bailey07} Bailey, J., Simpson, A.,
 Crisp, D., 2007, PASP, 119, 228.
\bibitem[\protect\citeauthoryear{Bailey et al.}{2008}]{bailey08} Bailey, J., Meadows, V.S., 
Chamberlain, S., Crisp, D., 2008, Icarus, 197, 247.
\bibitem[\protect\citeauthoryear{Banfield et al.}{1996}]{banfield96} Banfield, D., Gierasch, P.J.,
Squyres, S.W., Nicholson, P.D., Conrath, B.J., Matthews, K., 1996, Icarus, 121, 389.
\bibitem[\protect\citeauthoryear{Banfield et al.}{1998}]{banfield98} Banfield, D., Gierasch, P. J., 
Bell, M.,  Ustinov, E., Ingersoll, A. P., Vasavada, A. R., West, R. A. \& Belton, M. J. S., 1998, 
Icarus, 135, 230.
\bibitem[\protect\citeauthoryear{Bellucci et al.}{2004}]{bellucci04} Bellucci, G. et al., 2004, 
AdSpR, 34, 1640.
\bibitem[\protect\citeauthoryear{Borysow}{2002}]{borysow02} Borysow, A., 2002 A\&A, 390, 779.
\bibitem[\protect\citeauthoryear{Borysow et al.}{1989}]{borysow89} Borysow, A., Frommhold, L. 
\& Moraldi, M. 1989, ApJ, 336, 495.
\bibitem[\protect\citeauthoryear{Borysov et al.}{2002}]{borysov02} Borysov, A., Champion, J.P., 
Jorgensen, U.G. and Wenger, C., 2002, Mol. Phys., 100, 3583.
\bibitem[\protect\citeauthoryear{Brown}{2005}]{brown05} Brown, L.R., 2005, J. Quant. Spectrosc. 
Radiat. Trans., 96, 251.
\bibitem[\protect\citeauthoryear{Calvin et al.}{1995}]{calvin95} Calvin, W. M., Clark, R. N., 
Brown, R. H., Spencer, J. R., 1995, JGR, 100, E9, 19041
\bibitem[\protect\citeauthoryear{Campargue et al.}{2010a}]{campargue10a} Campargue, A., Wang, L., Liu, A.W.,
Hu, S.M., Kassi, S., 2010a, Chem. Phys., 373, 203.
\bibitem[\protect\citeauthoryear{Campargue et al.}{2010b}]{campargue10b} Campargue, A., Wang, L., Kassi, S,
Masat, M.,, Votava, O., 2010b, J. Quant. Spectrosc. Radiat. Trans., 111, 1141.
\bibitem[\protect\citeauthoryear{Carlson et al.}{1996}]{carlson96} Carlson, R. et al. 1996, Science,
274, 385.
\bibitem[\protect\citeauthoryear{Clark \& McCord}{1979}]{clark79} Clark, R. N. \& McCord, T. B. 
1979, Icarus, 40, 180.
\bibitem[\protect\citeauthoryear{Conrath \& Gierasch}{1986}]{conrath86} Conrath B. J. \& Gierasch, P.J.,
 1986, Icarus, 67, 444.
\bibitem[\protect\citeauthoryear{Conrath et al.}{1981}]{conrath81} Conrath, B. J., Flasar, F.M., 
Pirraglia, J.A., Gierasch, P.J. and Hunt, G.E. 1981, J. Geophys. Res., 86, 8769.
\bibitem[\protect\citeauthoryear{Conrath et al.}{1998}]{conrath98} Conrath, B.J., Gierasch, P.J. and Ustinov E.A. 1998, Icarus, 135, 501.
\bibitem[\protect\citeauthoryear{Frankenberg et al.}{2008}]{frankenberg08} Frankenberg, C., Warneke, T., Butz, A., Aben, I., Hase,
F., Spietz, P., Brown, L.R., 2008, Atmos. Chem. Phys., 8, 5061.
\bibitem[\protect\citeauthoryear{Folkner et al.}{1998}]{folkner98} Folkner, W.M., Woo, R. 
\& Nandi, S., 1998., J. Geophys. Res. 103,  22847.
\bibitem[\protect\citeauthoryear{Gao et al.}{2009}]{gao09} Gao, B., Kassi, S., Campargue, A., 2009, J. Mol. Spectrosc.,
253, 55.
\bibitem[\protect\citeauthoryear{Hilico et al.}{2001}]{hilico01} Hilico, J.-C., Robert, O., Loete, M., Toumi, S., Pine,
A.S., Brown, L.R., 2001, J. Mol. Spectrosc., 208, 1.
\bibitem[\protect\citeauthoryear{Irwin}{1999}]{irwin99a} Irwin, P. G. J., 1999, Surveys in Goephysics,
20, 505. 
\bibitem[\protect\citeauthoryear{Irwin et al.}{1999}]{irwin99b} Irwin, P.G.J., Calcutt, S.B., 
Sihra, K., Taylor, F.W., Weir, A.L., Ballard, J., Johnston, W.B., 1999, J. Quant. Spectrosc. 
Radiat. Trans., 62, 193.
\bibitem[\protect\citeauthoryear{Irwin et al.}{2005}]{irwin05} Irwin, P.G.J. Sihra, K., 
Bowles, N., Taylor, F.W., Calcutt, S.B., 2005, Icarus, 176, 255.
\bibitem[\protect\citeauthoryear{Karkoschka}{1994}]{karkoschka94} Karkoschka, E, 1994, Icarus, 
111, 174.
\bibitem[\protect\citeauthoryear{Kunde et al.}{1982}]{kunde82} Kunde, V. et al., 1982.  ApJ. 263,
443.
\bibitem[\protect\citeauthoryear{Kunde et al.}{2004}]{kunde04} Kunde, V. G., Flasar F.M. Jennings, D. E. et al.  2004, 
Science, 305, 1582.
\bibitem[\protect\citeauthoryear{Kurucz}{2009}]{kurucz09} Kurucz, R. L. "The Solar Irradiance by 
Computation", accessed on 12 Jan 2009 from {\it http://kurucz.harvard.edu/papers/irradiance/solarirr.tab}
\bibitem[\protect\citeauthoryear{Lindal et al.}{1981}]{lindal81} Lindal, G. F., Wood, G. E., 
Levy, G. S., Anderson, J. D., Sweetnam, D. N., Hotz, H. B., Buckles, B. J., holmes, D. P. \& Doms, P. E. 1981, 
JGR, 86, 8721.
\bibitem[\protect\citeauthoryear{Lindal}{1992}]{lindal92} Lindal, G. F., 1992, ApJ, 103, 967.
\bibitem[\protect\citeauthoryear{Liou}{2002}]{liou02} Liou, K, 2002, An Introduction to Atmospheric 
Radiation (2nd Edition), Academic Press, pp92--93.
\bibitem[\protect\citeauthoryear{Margolis et al.}{1990}]{margolis90} Margolis, J.S., 1990, Appl. 
Opt. 29, 2295.
\bibitem[\protect\citeauthoryear{Martonchik et al.}{1984}]{martonchik84} Martonchik, J. V., 
Orton, G. S. \& Appleby, J. F., 1984, Appl. Opt., 23, 541.
\bibitem[\protect\citeauthoryear{Mishchenko et al.}{2002}]{mishchenko02} Mishchenko, M. I., 
Travis, L. D. \& Lacis, A. A. 2002, {\it Scattering, Absorption and Emission of Light by Small Particles}, 
Cambridge University Press
\bibitem[\protect\citeauthoryear{Moses et al.}{2005}]{moses05} Moses, J. I., Fouchet, T., Bezard, B., Gladstone, G. R., 
Lellouch, E. \& Feuchtgruber, H. 2005, JGRE, J. Geophys. Res., 110, E08001.
\bibitem[\protect\citeauthoryear{Nikitin et al.}{2010}]{nikitin10} Nikitin, A.V. et al.,
2010, J. Quant. Spectrosc. Radiat. Trans., 111, 2211.
\bibitem[\protect\citeauthoryear{Pine}{1992}]{pine92} Pine, A.S., 1992, J. Chem. Phys., 97, 773.
\bibitem[\protect\citeauthoryear{Pine \& Gabard}{2003}]{pine03} Pine, A.S. \& Gabard, T., 2003, 
J. Mol. Spect., 217, 105.
\bibitem[\protect\citeauthoryear{Pryor \& Hord}{1991}]{pryor91} Pryor, W. R. \& Hord, C. W. 1991, 
Icarus, 91, 161.
\bibitem[\protect\citeauthoryear{Ragent et al.}{1998}]{ragent98} Ragent, B., Colburn, D. S., Rages, K. A., 
Knight, T. C. D., Avrin, P., Orton, G. S., Yanamandra-Fisher, P. A. \& Grams, G. W., 1998, JGR, 103,
22891.
\bibitem[\protect\citeauthoryear{Rothman et al.}{2009}]{rothman09} Rothman, L.S. et al., 
2009, J. Quant. Spectrosc. Radiat. Trans., 110, 533. 
\bibitem[\protect\citeauthoryear{Sault et al.}{2004}]{sault04} Sault, R.J., Engel, C. \& de Pater, 
I. 2004, Icarus, 168, 336.
\bibitem[\protect\citeauthoryear{Sciamma-O'Brien et al}{2009}]{sciamma09} Sciamma-O'Brien, E., Kassi, S.,
Gao, B., Campargue, A., 2009, J. Quant. Spectrosc. Radiat. Trans., 110, 951.
\bibitem[\protect\citeauthoryear{Seiff et al.}{1998}]{seiff98} Seiff, A. et al., 1998, JGR, 103, 22857.
\bibitem[\protect\citeauthoryear{Shortridge et al.}{1995}]{shortridge95} Shortridge, K., 
Meatheringham, S.J., Carter, B.D., Ashley, M.C.B., 1995, PASA, 12, 244.
\bibitem[\protect\citeauthoryear{Showman \& Dowling}{2000}]{showman00} Showman, A. P., Dowling, T. E. 2000, Science, 289, 1737.
\bibitem[\protect\citeauthoryear{Simon-Miller et al.}{2001}]{simon01} Simon-Miller, A. A., 
Banfield D. and Gierasch P. J. 2001, Icarus, 154, 459.
\bibitem[\protect\citeauthoryear{Simon-Miller et al.}{2006}]{simon06} Simon-Miller, A. A., 
Conrath, B. J., Gierasch P. J., Orton, G. S., Achtenberg, R. K., Flasar, F. M. and Fisher B. M. 2006, Icarus, 180, 98.
\bibitem[\protect\citeauthoryear{Stanmes et al.}{1988}]{stanmes88} Stanmes, K., Tsay, C. S., Wiscombe, W. \& Jayaweera, K. 
1988, Appl. Opt, 27(12), 2502.
\bibitem[\protect\citeauthoryear{Strobel}{1983}]{strobel83} Strobel, D. F., 1983, Int. Rev. Phys.
Chem., 3, 145.
\bibitem[\protect\citeauthoryear{Strong et al.}{1993}]{strong93} Strong, K., Taylor, F.W., 
Calcutt, S.B., Remedios, J.J., Ballard, J., 1993, J. Quant. Spectrosc. Radiat. Trans., 50, 363. 
\bibitem[\protect\citeauthoryear{Swain et al.}{2008}]{swain08} Swain, M.R., Vasisht, G., 
Tinetti, G., 2008, Nature, 452, 329.
\bibitem[\protect\citeauthoryear{Swain et al.}{2009}]{swain09} Swain, M.R. et al., ApJ, 704, 1616.
\bibitem[\protect\citeauthoryear{Tinney et al.}{2004}]{tinney04} Tinney, C.G. et al., 2004, 
Proc. SPIE, 5492, 998.
\bibitem[\protect\citeauthoryear{Tomasko et al.}{1978}]{tomasko78} Tomasko, M.G., West, R.A. \& 
Castillo, N. D. 1978, Icarus, 33, 558.
\bibitem[\protect\citeauthoryear{Tomasko et al.}{1986}]{tomasko86} Tomasko, M. G., Karkoschka, E. 
\& Martinek, S., 1986, Icarus, 65, 218.
\bibitem[\protect\citeauthoryear{Wang et al.}{2010}]{wang10} Wang, L., Kassi, S., Campargue,
A., J. Quant. Spectrosc. Radiat. Trans., 111, 1130.
\bibitem[\protect\citeauthoryear{Wenger \& Champion}{1998}]{wenger98}Wenger, C. \& Champion, J.P., 1998, JQSRT, 59, 471
\bibitem[\protect\citeauthoryear{West et al.}{1986}]{west86} West, R. A., Strobel, D. F. \& 
Tomasko, M. G., 1996, Icarus, 65, 161.
\bibitem[\protect\citeauthoryear{Wong et al.}{2004}]{wong04} Wong, M.H., Mahaffy, P.R., Atreya, S.K.,
Niemann, H.B \& Owen, T.C., 2004,  Icarus, 171, 153.
\bibitem[\protect\citeauthoryear{Yurchenko et al.}{2009}]{yurchenko09} Yurchenko, S.N., Barber, R.J., Yachmenev, A., Theil,
W., Jensen, P., Tennyson, J., 2009, J. Phys. Chem. A. 113, 11845.




\end{thebibliography}
\end{document}